\documentstyle [fleqn,12pt]{article}
\begin{document}
\baselineskip .75cm 
\begin{titlepage}
\title{\bf Comments on quasiparticle models of quark-gluon plasma}       
\author{Vishnu M. Bannur  \\
{\it Department of Physics}, \\  
{\it University of Calicut, Kerala-673 635, India.} }   
\maketitle
\begin{abstract}

Here we comment on the thermodynamic inconsistency problem and 
the reformulation of statistical mechanics of widely studied 
quasiparticle models of quark-gluon plasma. Their starting relation, 
the expression for pressure itself is a wrong choice and lead to 
thermodynamic inconsistency and the requirements of the 
reformulation of statistical mechanics. We propose a new approach to the problem using the standard statistical mechanics and is thermodynamically consistent. 
We also show that the other quasiparticle models may be obtained from 
our general formalism as a special case under certain restrictive condition. 
Further, as an example, we have applied our model to explain the nonideal 
behaviour of gluon plasma and obtained a remarkable good fit to the lattice 
results by adjusting just a single parameter.                                                      
\end{abstract}
\vspace{1cm}
                                                                                
\end{titlepage}
\section{Introduction :}

The quasiparticle quark-gluon plasma (qQGP) is a phenomenological model, 
with few fitting parameters, widely used to describe the nonideal behaviour 
of quark-gluon plasma (QGP). It was first proposed by Goloviznin and Satz 
\cite{sz.1} and then by Peshier {\it et. al.} 
\cite{pe.1} to explain the equation of state (EoS) of QGP from lattice gauge 
theory (LGT) simulation of quantum chromodynamics (QCD) at finite temperature 
\cite{l1.1}. The model, however, failed \cite{ba.1} to explain the more 
accurate, recent lattice data \cite{l2.1}. Further, Gorenstein and Yang 
\cite{go.1} pointed out that the model is thermodynamically inconsistent. 
This thermodynamically inconsistency problem was remedied by them by 
introducing a temperature dependent vacuum energy and 
forced it to cancel the thermodynamically   
inconsistent term, which was named as the reformulated statistical mechanics.
It is still not clear what is the physics or origin of this constraint which 
was called as thermodynamic consistency check in Ref. \cite{go.1,pe.2,lh.1,s.1} 
Here we show that the whole exercise is unnecessary and following the 
standard statistical mechanics (SM), we propose a new qQGP model which 
contains a single phenomenological parameter. Our model is thermodynamically 
consistent and explains lattice data very well. 
                                                                               
\section{Our model of qQGP:} 

Let us start with the work of Peshier {\it et. al.} \cite{pe.1} on gluon 
plasma. All thermodynamic quantities were derived from the pressure, 
$P$, which was assumed as 
\begin{equation} 
\frac{P\,V}{T} = - \sum_{k=0}^\infty \ln (1 -  
e^{ - \beta \epsilon_k})\,\, , \label{eq:p}  
\end{equation}  
where the right hand side is the logarithm of the grand partition function, 
$Q_G(T)$, and $\epsilon_k$ is the single particle energy of quasi-gluon, 
i.e, gluon with temperature dependent mass, given by,
\[ \epsilon_k = \sqrt{k^2 + m^2 (T)} \,\,, \]   
where $k$ is momentum and $m$ is mass. $\beta$ is defined as $1/T$. 
The expression for pressure 
is similar to that of ideal gas with temperature dependent mass and 
hence let us denote it as $P_{id}$. 
We want to stress that this assumption itself is 
the root cause of thermodynamic inconsistency and 
hence the reformulation of SM by Gorenstein and Yang \cite{go.1}. 
Generally, in grand canonical ensemble (GCE), energy ($E_r$) and number 
of particles ($N_s$) fluctuate, but temperature ($T$) and 
the chemical potential ($\mu$) are fixed. Hence, the average energy ($U$) 
and average number of particles ($N$) are defined and may be related to 
the grand partition function or q-potential, 
\begin{equation} 
q \equiv \ln Q_G = 
\ln ( \sum_{s,r} e^{- \beta E_r  - \alpha N_s}) 
= \mp \sum_{k=0}^\infty 
\ln (1 \mp z\, e^{ - \beta \epsilon_k})  
\,\,, \label{eq:q} \end{equation} 
where $\mp$ for bosons and fermions and  
$z \equiv e^{\mu /T} = e^{-\alpha}$ is called fugacity. 
The average energy $U$ is defined as,   
\begin{equation}
U \equiv <E_r> = \frac{\sum_{s,r}\,E_r \,e^{- \beta E_r - \alpha N_s}}{Q_G} = 
- \frac{\partial }{\partial \beta} \ln Q_G = 
\sum_k \frac{z \,\epsilon_k e^{- \beta \epsilon_k }}{1 \mp 
z\, e^{- \beta \epsilon_k} } 
\,\,. \label{eq:uo} 
\end{equation}
Note that the partial differentiation with repect to $\beta$ above is just 
a mathematical method to express $U$ in terms of sum over single particle 
energy levels, $\epsilon_k$, making use of Eq. (\ref{eq:q}). While 
differentiating, indirect dependence of $\beta = 1/T$ in the fugacity, $z$, 
and mass, $m(T)$, must be ignored by definition. 
Otherwise, we will not get back $<E_r>$.  
Similarly, the average density $N$ is defined as, 
\begin{equation}
N \equiv <N_s> = \frac{\sum_{s,r}\,N_s \,e^{- \beta E_r - \alpha N_s}}{Q_G} = 
- \frac{\partial }{\partial \alpha} \ln Q_G = 
 z \frac{\partial }{\partial z} \ln Q_G =  
\sum_k \frac{z \,e^{- \beta \epsilon_k }}{1 \mp 
z\, e^{- \beta \epsilon_k} } \,\,,\label{eq:n}
\end{equation}  
These (Eqs. (\ref{eq:uo}), (\ref{eq:n})) are the standard relations \cite{pa.1} 
of $U$ and $N$ to the partition function, 
which is valid even for quasiparticle with ($T$,$\mu$) dependent mass 
by the definition of averages. 
Here, for gluon plasma, we have $\mu = 0$ or $z=1$.  
Next, pressure may be obtained by two methods. In method-I, one starts from 
$U$ and using thermodynamic relation, 
\begin{equation}
\varepsilon \equiv \frac{U}{V} =  T \frac{\partial P}{\partial T} 
- P \,\, ,  \label{eq:td} 
\end{equation}
and on integration, one gets pressure which is the procedure that we follow here. 
In method II, again following the standard text books on SM \cite{pa.1}, 
we can relate $P$ to q-potential as follows. The variation 
in q-potential due to variations in it's dependence, namely  
$T$, $\mu$ and volume $V$, specifying the macro-state of GCE system, is,   
\begin{equation}
\delta q = \frac{1}{Q_G} \left[ \sum_{r,s} e^{- \beta (E_r - \mu N_s)}
\, \left( - E_r \,\delta \beta - \beta \,\delta E_r + N_s \,\delta (\beta \mu) 
\,\right) \right] \,\,.
\end{equation}
Now, when compared with the text books results, we have an extra term coming 
from $\delta E_r$ due to temperature dependent mass  
and then using the definition of averages, we get,  
\begin{equation} 
 \frac{P\,V}{T} = \mp \sum_{k=0}^\infty \ln (1 \mp z\,
e^{ - \beta \epsilon_k}) + \int d\beta\, \beta \, 
\frac{\partial m}{\partial \beta}\,<\frac{\partial E_r}{\partial m}> 
\,\,. \label{eq:pn}  
\end{equation}  
Therefore we see that $P$ is not just equal to $P_{id}$, but there is an 
extra term. This extra term ensure thermodynamic consistency of the relation 
as follows. From above $P$, on differentiating with respect to $T$ for 
a system with $\mu=0$ or $z=1$,  
\begin{equation}
\frac{\partial P}{\partial T} = \frac{P}{T} + \frac{\varepsilon}{T} 
- \frac{1}{V}\,< \frac{\partial \epsilon_k}{\partial T} >  
+ \frac{1}{V}\,< \frac{\partial E_r}{\partial T} > 
\end{equation}
where the last two terms exactly cancels (following the procedure used 
in deriving Eq, (\ref{eq:uo})) and hence the thermodynamic 
relation, Eq. (\ref{eq:td}), is obeyed as expected.   

Further more, this $P$ is also consistent with the $P$ obtained from $U$ 
through thermodynamic relations which may be shown as follows. 
The Eq. (\ref{eq:pn}) may be simplified by evaluating 
$<\frac{\partial E_r}{\partial m}>$ and taking continuum limit 
($V \rightarrow \infty$), for a system with $\mu = 0$, as, 
\begin{equation} 
 \frac{P}{T} = \mp \frac{g_f}{2 \pi^2} \int_0^{\infty} dk\,k^2\,
\ln(1 \mp \,e^{-\beta \epsilon_k})  + \int d\beta \,\beta  
\frac{g_f}{2\pi^2}\,m\,\frac{dm}{d\beta}\,\int_0^{\infty} dk 
\frac{k^2}{\epsilon_k \,( e^{\beta \epsilon_k} \mp 1)} 
\,\,, \label{eq:pn1} \end{equation} 
which on simplification, reduces to, 
\[ \frac{P}{T} = \frac{g_f}{2\,\pi^2} \left[ T^3\, \sum_{l=1}^{\infty} 
(\pm 1)^{l-1}\,\frac{1}{l^4}\, (\beta\,m\,l)^2\,K_2 (\beta\,m\,l)\, \right. \]
\begin{equation} 
\left. +\,  \int d\beta \frac{\beta}{m}\,\frac{\partial m}{\partial \beta}\,
\frac{1}{\beta^4} \sum_{l=1}^{\infty} 
(\pm 1)^{l-1}\,\frac{1}{l^4}\, (\beta\,m\,l)^3\,K_1 (\beta\,m\,l)\,\right]   
\,\,, \end{equation} 
where $g_f$ is the internal degrees of freedom and $K_1$, $K_2$ are 
modified Bessel functions. Using the recurrence relations of 
Bessel functions and on integration by parts, above equation may 
be further simplified to get,
\begin{equation} 
\frac{P}{T} = \frac{P_0}{T_0} - \int_{\beta_0}^{\beta} \, d\beta\, 
\varepsilon \,\,, \label{eq:td2} \end{equation} 
where $\varepsilon$ is the energy density and $P_0$ is the pressure 
at some temperature $T_0$ or $\beta_0$. This equation is nothing 
but the thermodynamic relation, Eq. (\ref{eq:td}).  
Therefore, Gorenstein and Yang's starting argument that above two 
methods give different $m(T)$ does not exist now by using our 
derived expression for $P$, instead of the assumption \cite{pe.1,go.1}. 

\section{Question of vacuum energy $B(T)$ :}

After the reformulation of SM by 
Gorenstein and Yang, almost all study in qQGP is based on the    
thermodynamic consistency relation, related to vacuum energy $B(T)$. Different authors call and interpret $B(T)$ in a different way, like vacuum energy, background field or bag pressure.  But, by definition of quasiparticle, whole thermal energy 
is used to excite these quasiparticles. So quasiparticles are excitations 
above the ground state or vacuum state which may not depend on 
temperature or chemical potential. This is our assumption. As noted earlier, we also don't have any thermodynamic inconsistency in our model. 

In fact, when we redo our derivation of pressure   
with vacuum or zero point energy in single particle energy,   
like in Ref. \cite{go.1}, Eq. (\ref{eq:pn1}) is modified as,  
\begin{equation} 
P = P_{id} - B(T) + T\,\left( \int_{T_0}^T \frac{d\tau}{\tau} 
\left[ \frac{g_f}{2\pi^2}\,m\,\frac{dm}{d\tau}\,\int_0^{\infty} dk 
\frac{k^2}{\epsilon_k \,(e^{\epsilon_k/\tau} \mp 1)} 
+ \frac{\partial B}{\partial \tau} \right] \right) \,\,, \label{eq:pn2}   
\end{equation}  
and the energy density, 
\begin{equation}
\varepsilon = \varepsilon_{id} + B(T) \,\,. 
\end{equation}
where $\varepsilon_{id}$ is the expression for $\varepsilon$ similar to 
ideal gas. Again it is easy to show that above $P$ and $\varepsilon$ 
obey thermodynamic relation Eq. (\ref{eq:td}).  
The thermodynamic consistency relation \cite{go.1}, used in other qQGP 
models, is nothing but a restrictive condition that the terms inside 
the square bracket in Eq. (\ref{eq:pn2}) is zero. At present it is 
not clear what is the physical origin of this constraint. Note that     
without this constraint, so called thermodynamic consistency relation, 
our system is thermodynamically consistent even with the zero-point energy contribution, $B(T)$. One may model $B(T)$ based on other effects of strongly interacting QCD system, like hadronic states, resonances and may be relevent at the 
transition point. In our study of gluon plasma here, we neglect all 
these effects and consider only the thermal properties of gluons. Hence 
we take $B(T) = 0$ and we get a very good fit to lattice results except at very close to the transition temperature, i.e, for $T<1.2 T_c$.        
    
\section{EoS of gluon plasma:} 

As an example, let us apply our model to gluon plasma which is a QCD plasma 
without quarks. We first calculate 
the energy density, expressed in terms of 
$e(T) \equiv \varepsilon / \varepsilon_s$,    
and then obtain $P$ from thermodynamic relation, Eq. (\ref{eq:td}). So we have, 
from Eq. (\ref{eq:uo}) after some algebra, 
\begin{equation}
e(T) = \frac{15}{\pi ^4} 
\sum_{l=1}^\infty  \frac{1}{l^4} 
\left[ (\frac{m_g\,l}{T})^3 K_1 (\frac{m_g\,l}{T}) 
+  3\, (\frac{m_g\,l}{T})^2 K_2 (\frac{m_g\,l}{T}) \right] 
\end{equation}
where $\varepsilon_s$ is the Stefan-Boltzman gas limit of QGP, 
$m_g$ is the temperature dependent mass and $K_1$, $K_2$ are 
modified Bessel functions.    
The results are plotted in Fig. 1, for two different mass 
terms, $m_g^2(T) = \omega_p^2 =  g^2(T)\,T^2\,/3$ (our model) 
and  $m_g^2(T) =  g^2(T)\,T^2\,/2$ (other qQGP models).   
$g^2(T)$ is related to the two-loop order running coupling constant, given by, 
\begin{equation} \alpha_s (T) = \frac{6 \pi}{(33-2 n_f) \ln (T/\Lambda_T)}
\left( 1 - \frac{3 (153 - 19 n_f)}{(33 - 2 n_f)^2}
\frac{\ln (2 \ln (T/\Lambda_T))}{\ln (T/\Lambda_T)}
\right)  \label{eq:ls} \;, \end{equation}
where $\Lambda_T$ is a parameter related to QCD scale parameter. This 
choice of $\alpha_s (T)$ is an approximate expression of the running coupling 
constant used in lattice simulations \cite{l2.1}. 
Then the pressure is obtained from the 
thermodynamic relation Eq. (\ref{eq:td}) or Eq. (\ref{eq:td2}).  
Since we have only one parameter to adjust, we don't get good fit for 
the generally used second choice of quasi-gluon mass. The best fitted 
parameter is $\Lambda_T /T_c = 0.3$. But a remarkably good 
fit may be obtained for our choice of gluon mass which is motivated from the 
fact that the quasi-gluons are the thermal excitations of plasma waves 
with mass equal to the plasma frequency \cite{me.1}. The value of the fitted 
parameter is $\Lambda_T /T_c = 0.65$.    

Let us now compare our results with the results from other qQGP models, for example Ref. \cite{pe.2}, where $B(T)$ is not zero, but is determined by thermodynamic consistency relation. From the Fig. 2, we see that only at large $T/T_c$ both the results almost match, but differ near to $T/T_c = 1$. We used the same $\alpha_s(T)$ with $\Lambda_T/T_c = .65$ for both the cases. 
Further, results from our model with $B(T) = 0$ 
fits well the lattice data. A very good fit to lattice data was 
also obtained in Ref. \cite{pe.2}, but with a different expression for 
$\alpha_s(T)$, having two free parameters, and an additional  parameter related to degrees of freedom.          

\section{Conclusions:} 

Here we have pointed out the basic reason for the thermodynamic inconsistency 
of the extensively studied quasiparticle QGP models \cite{pe.1} and it's consequence of the reformulation of statistical mechanics \cite{go.1}. 
To revise it, we have proposed a new qQGP model which follows from the standard SM and has no thermodynamic inconsistency. When we extend our formalism to a system with temperature dependent vacuum energy, again, we get a thermodynamically consistent general model and we obtained other widely studied qQGP models as a special case of our model under certain restrictive condition which was called as 
thermodynamic consistency relation in Ref. \cite{go.1,pe.2,lh.1,s.1}. 
As an example, we studied the gluon plasma 
using our model. A remarkable good fit to the 
LGT data was obtained by adjusting just one parameter and without 
the temperature dependent vacuum energy $B(T)$. Whereas we know that the 
other qQGP models has 3 or more parameters. Further extension of our 
model to flavored QGP without and with masses, and also without and with 
chemical potential, fit remarkably well 
the lattice results and were reported in \cite{ba.2,ba.3}.

\newpage
\begin{figure}
\caption { Plots of $P/ T^4 $ (lower set of graphs) and  
$\varepsilon/ T^4$ (upper set of graphs)  as a function of $T/T_c$ from 
our model and lattice results (symbols) \cite{l2.1} for gluon plasma with 
two different models for mass, $m_g^2(T) = g^2 T^2 /3$ (dashed line) 
and $m_g^2(T) = g^2 T^2 /2$ (dashed-dotted line).} 
\label{fig 1}
\vspace{.75cm}
\caption { Plots of $P/ T^4 $ (lower set of graphs) and  
$\varepsilon/ T^4$ (upper set of graphs)  as a function of $T/T_c$ from 
our model and lattice results (symbols) \cite{l2.1} for gluon plasma with 
$B(T)=0$ (dashed line) and using thermodynamic consistency relation 
(dashed-dotted line).} 
\label{fig 2}
\vspace{.75cm}
\end{figure}
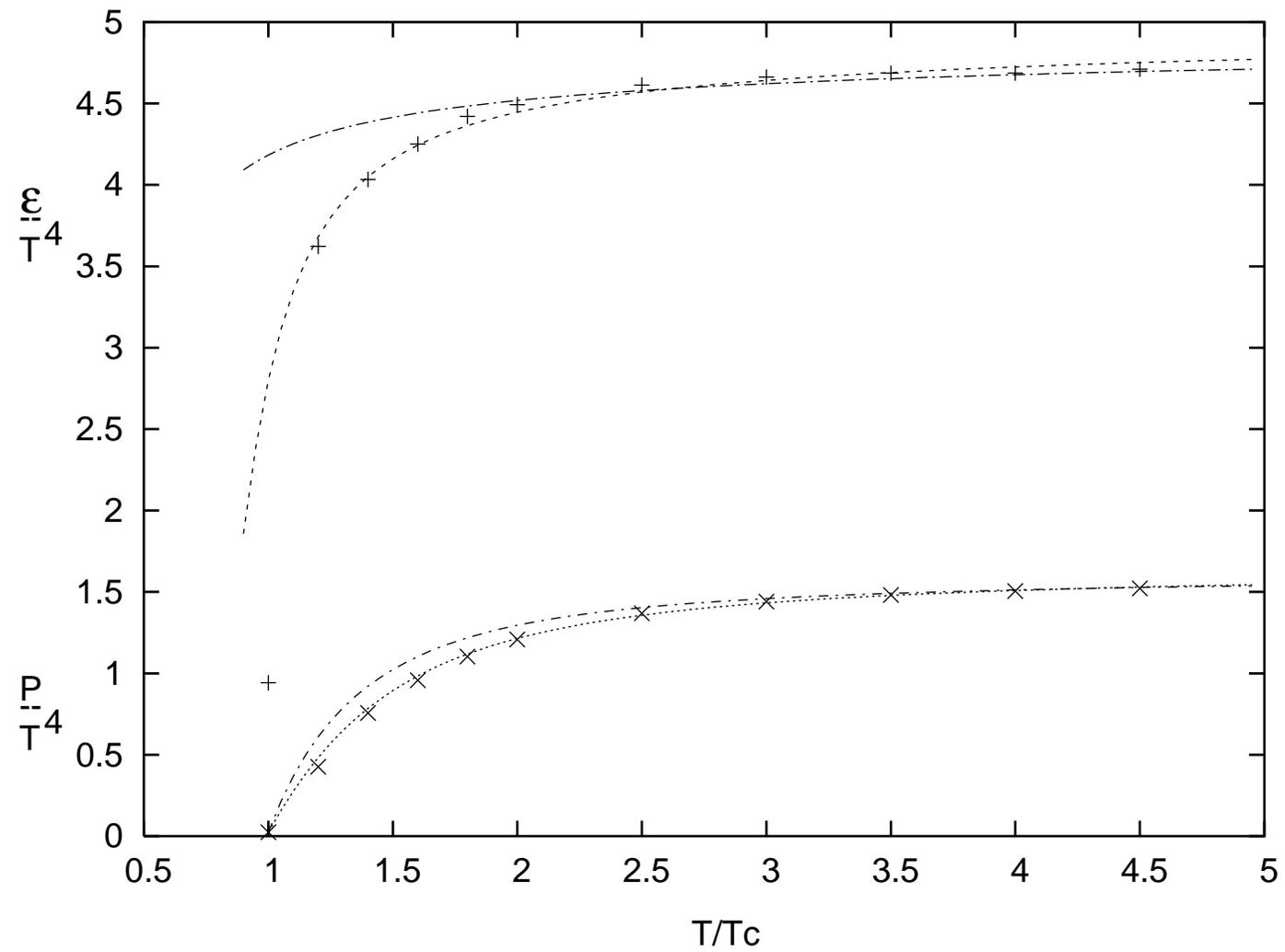
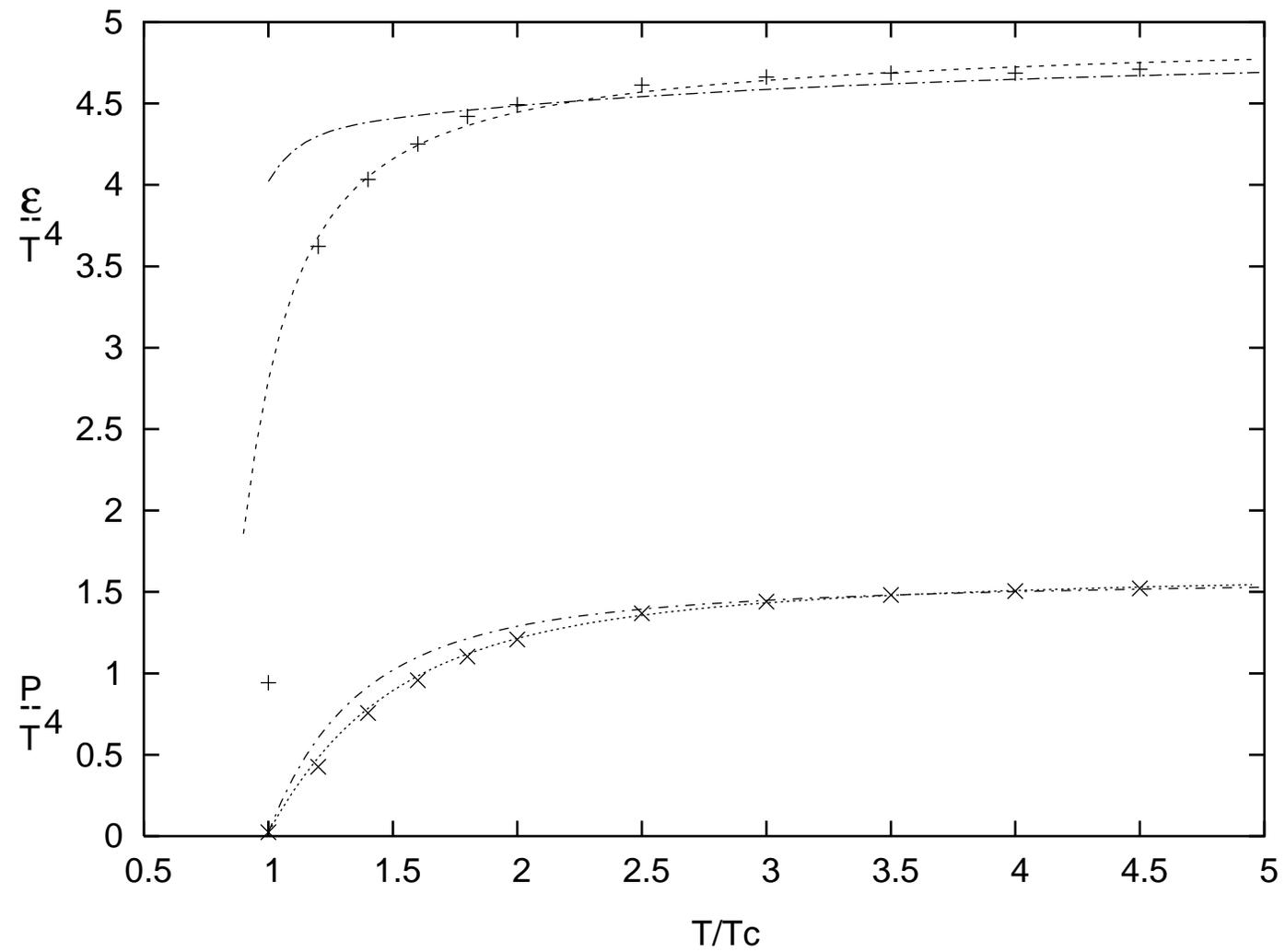
\end{document}